\documentclass[10pt,american,english]{article}
\usepackage[T1]{fontenc}
\usepackage[latin9]{inputenc}
\usepackage[letterpaper]{geometry}
\usepackage{float}
\usepackage{textcomp}
\usepackage{amssymb}
\usepackage{graphicx}
\usepackage{setspace}
\usepackage[numbers]{natbib}

\makeatletter

\usepackage{multicol}
\usepackage{fancyhdr}
\usepackage{titlesec}
\usepackage[flushleft]{threeparttable}
\usepackage{caption}
\usepackage{float}



\titleformat*{\section}{\normalsize\bfseries}
\titleformat*{\subsection}{\normalsize\itshape}
\titleformat*{\subsubsection}{\normalsize\itshape}
\titleformat*{\paragraph}{\normalsize\itshape}
\titleformat*{\subparagraph}{\normalsize\itshape}

\makeatother

\usepackage{babel}
\begin{document}
\begin{center}
IAC-16-C2.6.6.32493
\par\end{center}

\medskip{}

\begin{center}
\textbf{TEMPERATURE RESTRICTIONS FOR MATERIALS USED IN AEROSPACE INDUSTRY
FOR THE NEAR-SUN ORBITS}
\par\end{center}

\medskip{}

\begin{center}
\textbf{Elena Ancona}\textsuperscript{a} and \textbf{Roman Ya. Kezerashvili}\textsuperscript{b}
\par\end{center}

\begin{center}
\textsuperscript{a}Politecnico di Torino, Corso Duca degli Abruzzi
24, TO, 10129, Torino, Italy, elena.ancona@gmail.com
\par\end{center}

\begin{center}
\textsuperscript{b}Department of Physics, New York City College of
Technology, The City University of New York, 300 Jay Street, Brooklyn
NY, 11201, New York City, USA, rkezerashvili@citytech.cuny.edu
\par\end{center}

\medskip{}

\begin{center}
\textbf{Abstract}
\par\end{center}

For near-Sun missions, the spacecraft approaches very close to the
Sun and space environmental effects become relevant. Strong restrictions
on how much close it can get derive from the maximum temperature that
the used materials can stand, in order not to compromise the spacecraft's
activity and functionalities. In other words, the minimum perihelion
distance of a given mission can be determined based on the materials'
temperature restrictions. The temperature of an object in space depends
on its optical properties: reflectivity, absorptivity, transmissivity,
and emissivity. Usually, it is considered as an approximation that
the optical properties of materials are constant. However, emissivity
depends on temperature. The consideration of the temperature dependence
of emissivity and conductivity of materials used in the aerospace
industry leads to the conclusion that the temperature dependence on
the heliocentric distance is different from the case of constant optical
properties \citep{Rom_desorp}. Particularly, taking into account
that emissivity is directly proportional to the temperature, the temperature
of an object increases as $r{}^{-2/5}$ when the heliocentric distance
$r$ decreases. This means that the same temperature will actually
be reached at a different distance and, eventually, the spacecraft
will be allowed to approach closer to the Sun without compromising
its activities. We focused on metals used for aerospace structures
(Al, Ti), however our analysis can be extended to all kinds of composite
materials, once their optical properties - in particular emissivity
- are defined.

\smallskip{}

\textbf{Keywords}: Near-Sun orbits, temperature restrictions for materials.

\medskip{}

\begin{multicols}{2}

\section{Introduction}

Materials in space are nowadays a key topic of research for every
aerospace industry. In fact, their characteristics and performances
are extremely relevant for every feature of the mission, including
the cost: the main objective is to obtain very light materials with
great mechanical properties. Aluminum has been for long time a valid
compromise for its weight and good characteristics, however the actual
trend is to shift to carbon fiber and composite structures, and possibly
in the future to combinations of plastics and various hybrid materials
such as metal-matrix composites, that will greatly reduce the weight
of a spacecraft and also its launch costs. Indeed, carbon fiber has
already replaced many spacecraft components, except for bulkheads
that are still made from titanium, aluminum or other conventional
metals and alloys because of the tremendous thermal and mechanical
demands. All materials have a well defined range of temperature in
which their behavior is optimal. Out of these bounds, their properties'
degradation leads to failing performances. Temperature is one of the
most considerable drivers, especially for Near-Sun missions, with
electromagnetic radiation and solar wind. When approaching the Sun,
the increasing temperature could become unbearable, depending on the
specific material characteristics. Moreover, the temperature reached
by the spacecraft depends not only on the outer environment, but also
on the spacecraft itself. In particular, the spacecraft temperature
is function of its materials' optical properties, that in turn are
function of temperature. It is clear, then, that a better knowledge
of the materials' optical properties dependence on temperature could
be the key to design properly a Near-Sun mission. Of course, consideration
of optical degradation of materials due to proton and electron radiation
for a Near-Sun mission is very important, however we do not address
this issue in the present study. 

This paper is organized in the following way: in Section 2 a brief
overview of the most commonly used materials in space applications
is given, while Section 3 provides the key features of temperature
dependence of a spacecraft material on heliocentric distance. Our
analysis and results for metals are reported in Section 4, whereas
Section 5 suggests plausible future developments of the study. Conclusions
follow in Section 6.

\section{Materials in space}

The principal design driver for a spacecraft is weight: the main challenge
is to minimize mass, and consequently launch costs, without compromising
reliability and functionality. Moreover, a spacecraft must accommodate
the payload and its subsystems, satisfying the mounting requirements,
and then support itself and its payload through all phases of the
mission, included the launch. In particular, not only good stiffness
and strength properties are required, but also oscillation and resonance
frequencies of structures must be taken into account. Hence, the design
of spacecraft structures needs an extremely careful selection of materials
based upon their strength, stiffness, damage tolerance, thermal and
electrical properties, as well as corrosion resistance and shielding
capabilities. For this reason, the space sector has traditionally
been a promoter for the development and the application of advanced
engineering materials. Our study has been conducted for the two most
vastly used metals in aerospace industry: in fact, considering their
strength and density, it is clear why aluminum and titanium are the
preferred for lightweight aerospace alloys. Titanium alloys are used
where lighter aluminum alloys no longer meet strength, corrosion resistance
and elevated temperature requirements \citep{MatSP}, as aluminum
has a fusion temperature of 660 \textdegree C, whereas titanium of
1668 \textdegree C. Aluminum has a low density, good specific strength
and it is easily workable. For its range of applicability it is also
cheap and widely available. However, its weak spot is the low melting
point. Titanium is more expansive but guarantees better performance
in hostile environments due to its great corrosion resistance. Moreover,
it can bear high temperatures.

\section{Temperature dependence on heliocentric distance}

It might be convenient to remark the difference between environmental
temperature in space and the temperature of an object, such as a spacecraft,
in space.

\subsection{Environmental temperature in space}

Matter in space is extremely concentrated into celestial bodies. The
space between them could be considered as a near-vacuum, where particles
may be many miles apart. Hence, under outer space conditions, almost
no energy is transferred ``directly'' because of the vast distances
involved. As a result, radiation is effectively the only heat-exchanging
mechanism in most of the space environment. In fact, the average temperature
of interplanetary space is 2.7 K, due to the cosmic microwave background.

\subsection{Spacecraft temperature dependence on heliocentric distance}

An object in space radiates heat and receives heat radiated from other
bodies. The result of this energy balance is that if the considered
body radiates more heat than it receives it will cool down; on the
contrary, if the incoming heat is more than that radiated, it will
warm up. As the intensity of received radiation decreases with the
square of the distance from the energy source, a spacecraft approaching
the Sun will be exposed to higher level of electromagnetic radiation.
Distance from stars and radiation exposure are the prime temperature
determinants for an object in space. In fact, if one side of a spacecraft
is exposed to direct sunlight and radiation, while the other side
is shadowed and facing out into deep space, the spacecraft would suffer
an extreme temperature differential which, if not sapiently contrasted,
could be dangerous and critical for the system survival.

It is clear then that the temperature of a spacecraft depends on its
optical properties: reflectivity, absorptivity, transmissivity and
emissivity. In fact, the solar electromagnetic radiation can be reflected,
absorbed or transmitted. Therefore, one can write:

\begin{equation}
\rho\left(\lambda,\mathit{T}\right)+\alpha\left(\lambda,T\right)+\tau\left(\lambda,T\right)=1,
\end{equation}

\noindent where $\rho\left(\lambda,T\right)$, $\alpha\left(\lambda,T\right)$
and $\tau\left(\lambda,T\right)$ are the radiative or optical properties
of the material: spectral (as they depend on wavelength $\lambda$)
hemispherical (as they are not directional) reflectivity, absorptivity
and transmissivity, respectively \citep{heatmass}. The contribution
of transmissivity $\tau$ can be neglected, as experimental data confirm
it is a very small fraction (almost 2\%) of the incoming solar energy
flux. Indeed, once that energy has been absorbed, it can also be emitted
from the surface, as a secondary process. From the Stefan-Boltzmann's
law, the rate of energy emitted from a surface at a certain temperature
is proportional to the fourth power of the temperature.

In order to estimate the temperature of a body's surface its effective
temperature is commonly used. The effective temperature of a generic
object is defined as the temperature of a black body that would emit
the same total amount of electromagnetic radiation \citep{ClarkeAstr}:

\begin{figure}[H]
\noindent \begin{centering}
\includegraphics[width=9cm]{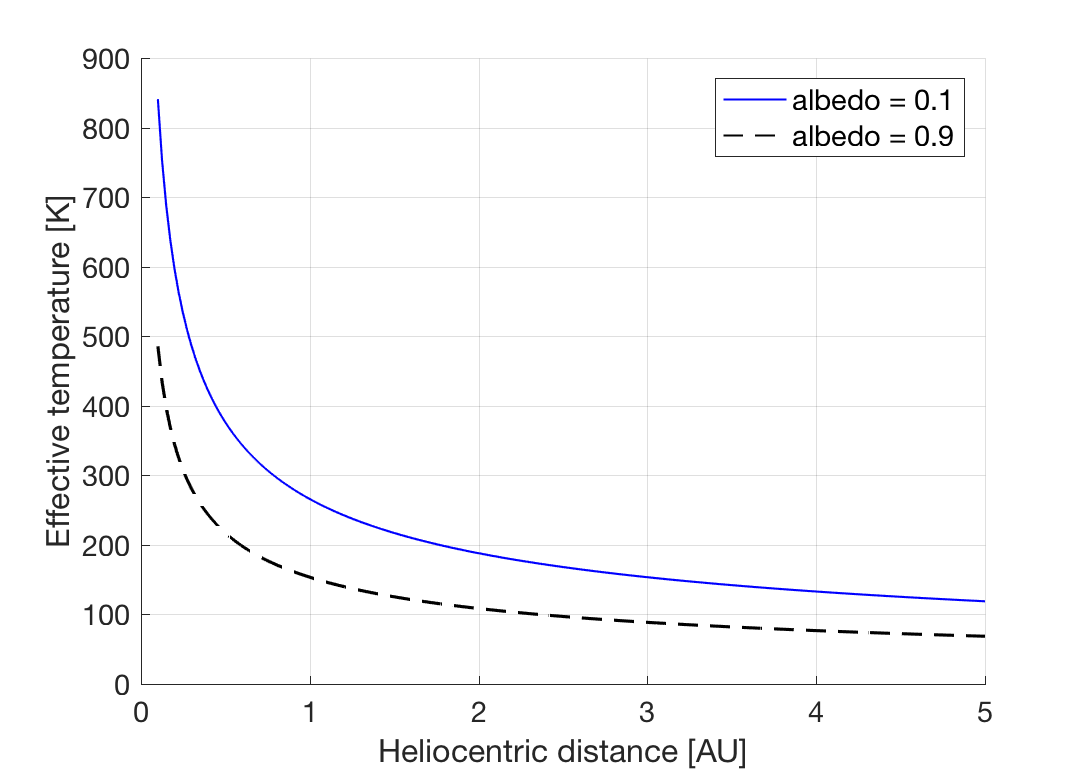}
\par\end{centering}

\caption{\textsl{Temperature dependence on reflectivity. The graph shows the
effective temperature of a body with different values of reflectivity
(albedo). It is the temperature of a black body ($\varepsilon=1$)
that would emit the same total amount of electromagnetic radiation.
If the body has lower emissivity, its actual temperature will be higher
of $T_{eff}$. The solid curve represents a body with low reflectivity
$a=0.1$, whereas the dashed curve a body with high reflectivity $a=0.9$.
The effective temperature of a body with generic albedo would be between
these limits.}}

\label{T_eff}
\end{figure}

\begin{equation}
T_{eff}=\left(\frac{L\left(1-a\right)}{16\pi\sigma r^{2}}\right)^{1/4},\label{eq:Teff}
\end{equation}

\noindent where $L$ is the star's luminosity, $a$ is the albedo,
$\sigma=5.67\cdot10^{-8}Wm^{-2}K^{-4}$ is the Stefan-Boltzmann constant
and $r$ is the distance of the object from the star, all in SI units.
The luminosity of a star depends on its radius $R_{S}$ and surface
temperature $T_{S}$:

\begin{equation}
L=4\pi\sigma R_{S}^{2}T_{S}^{4}.
\end{equation}

\begin{figure}[H]
\noindent \begin{centering}
\includegraphics[width=9cm]{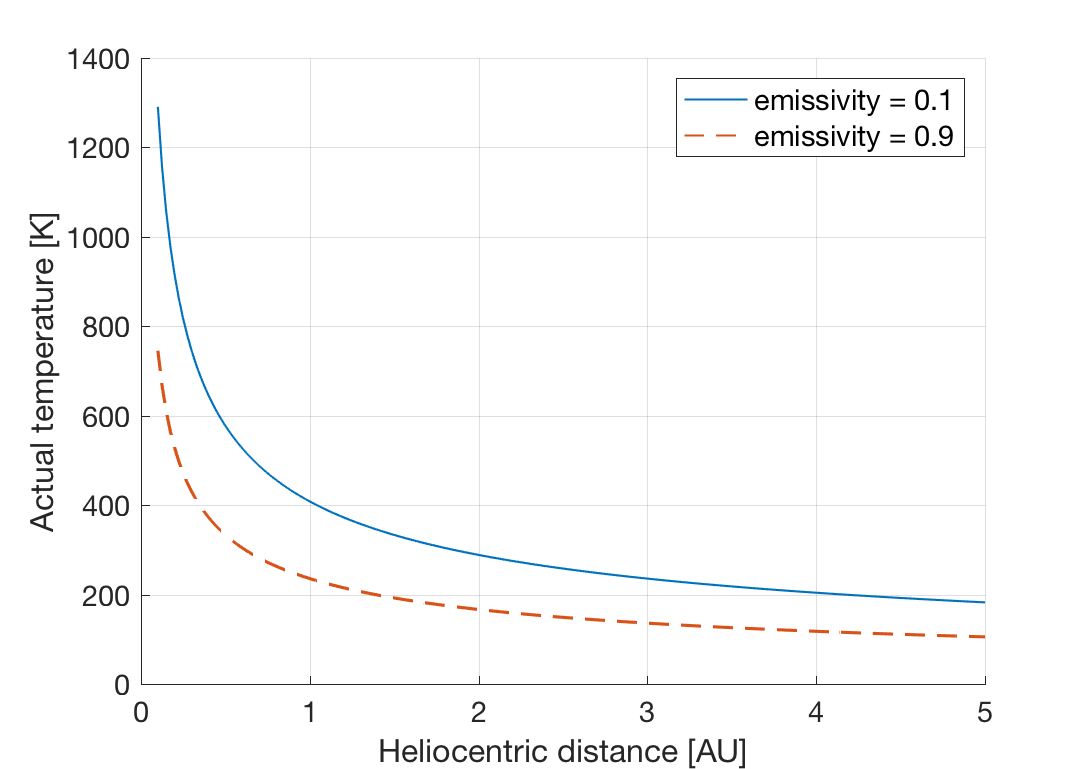}
\par\end{centering}

\caption{\textsl{Temperature dependence on emissivity. The graph shows the
actual temperature of a generic body with defined albedo $a=0.5$,
for different values of emissivity. The solid curve represents a body
with low emissivity $\varepsilon=0.1$, whereas the dashed curve a
body with high emissivity $\varepsilon=0.9$. The actual temperature
of a body with generic emissivity $0.1<\varepsilon<0.9$ would be
between these limits.}}

\label{T_act}
\end{figure}

For what concerns the albedo, it is a reflection coefficient, defined
as the ratio of radiation reflected from the surface to the incident
radiation, so $a=0$ for a perfectly black surface and $a=1$ for
a perfectly white surface\footnote{The albedo changes in the range 0 - 1 depending on the surface material,
e.g. $a=0.06$ for the Moon, which main surface component is basalt,
$a=0.16$ for Mars (iron oxide), $a=0.38$ for Earth, and $a=0.70$
for Jupiter (gas). }. Fig. \ref{T_eff} shows the effective temperature dependence on
heliocentric distance for different values of albedo. Eq. (\ref{eq:Teff})
is useful when the real emissivity of the specific body is unknown.
However, it can be modified to account for the actual emissivity of
a body $\varepsilon$, leading to more accurate results:

\begin{equation}
T_{act}=\left(\frac{A_{abs}}{A_{rad}}\frac{L\left(1-a\right)}{4\pi\sigma\varepsilon r^{2}}\right)^{1/4},\label{eq:Tact}
\end{equation}
where $A_{abs}$ and $A_{rad}$ are the portion of the total area
involved into absorption and radiation, respectively.

Eq. (\ref{eq:Tact}) allows to evaluate the surface temperature of
celestial bodies. The ratio $A_{abs}/A_{rad}$ is assumed to be 1/4
for a rapidly rotating body and 1/2 for a slowly rotating body. Note
that the net emissivity of the planet may be lower due to surface
or atmospheric properties, such as the greenhouse effect. When the
planet's net emissivity in the relevant wavelength band is less than
unity (less than that of a black body), the actual temperature of
the body will be higher than the effective temperature: $T_{act}>T_{eff}$.
In other words, $T_{eff}$ is the lower limit for the temperature
of a body with a given reflectivity ratio, at a distance $r$ from
a star of luminosity $L$. The actual temperature of a body with a
mean value of reflectivity coefficient ($a=0.5$) is shown in Fig.
\ref{T_act}, for two different values of emissivity.

Eq. (\ref{eq:Tact}) is sometimes written in a more functional form\footnote{http://spacemath.gsfc.nasa.gov.}:

\begin{equation}
T_{act}=273\left(\frac{\mathfrak{L}\left(1-a\right)}{\varepsilon\mathsf{r}^{2}}\right)^{1/4},\label{eq:Tact-1}
\end{equation}
where $\mathfrak{L}=L/L_{SUN}$ is the luminosity of a generic star
in multiples of the Sun's power and $\mathsf{r}$ is the distance
between the body and the star in AU. As a result, the temperature
of an object increases as $r{}^{-1/2}$ when the heliocentric distance
decreases, if all the other parameters are constant.

\section{Temperature dependence on heliocentric distance for metals}

Let us focus our attention on metals, largely used in the aerospace
industry.

In Section 3 a method for evaluating the surface temperature of a
body has been provided: Eq. (\ref{eq:Tact}) can be applied to any
material, once its constant emissivity and reflectivity are known.
It requires as input the emissivity and reflectivity of the material,
and gives as result its surface temperature. However recent studies
show that reflectivity and emissivity also depend on temperature \citep{materials}.

\begin{figure}[H]
\noindent \begin{centering}
\includegraphics[angle=-0.4,width=8cm]{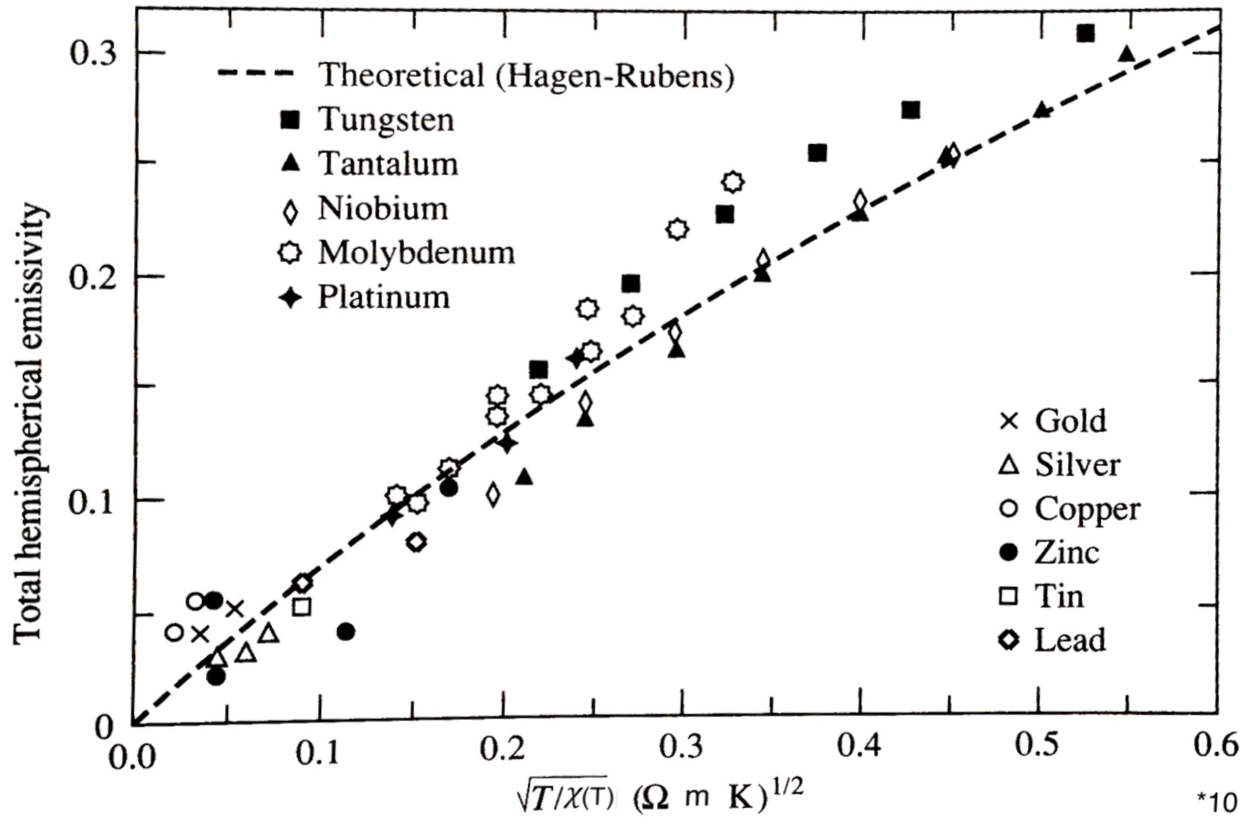}
\par\end{centering}

\caption{\textsl{Total hemispherical emissivity of various metals as function
of $\sqrt{T/\chi\left(T\right)}$, where $\chi$ is the material's
conductivity \citep{radheat}.}}

\label{emissivity}
\end{figure}

\begin{figure*}[t!]

\begin{minipage}[c]{\textwidth} \centering 

\selectlanguage{american}%
\includegraphics[width=9cm]{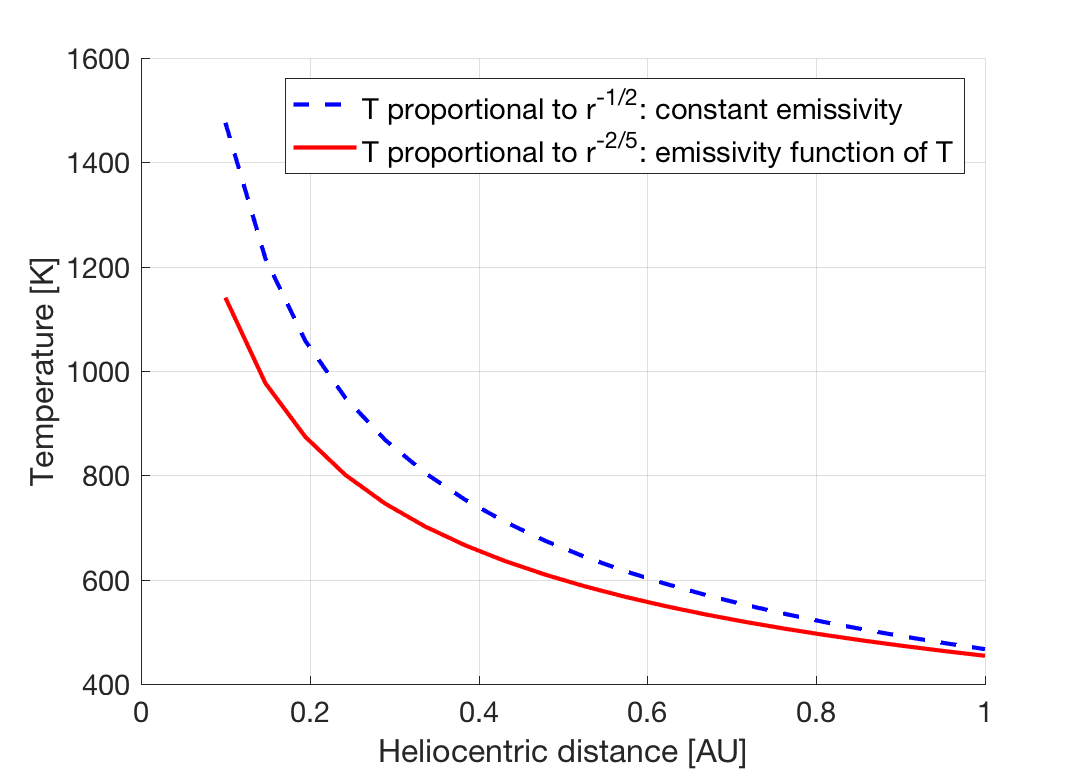}\includegraphics[width=9cm]{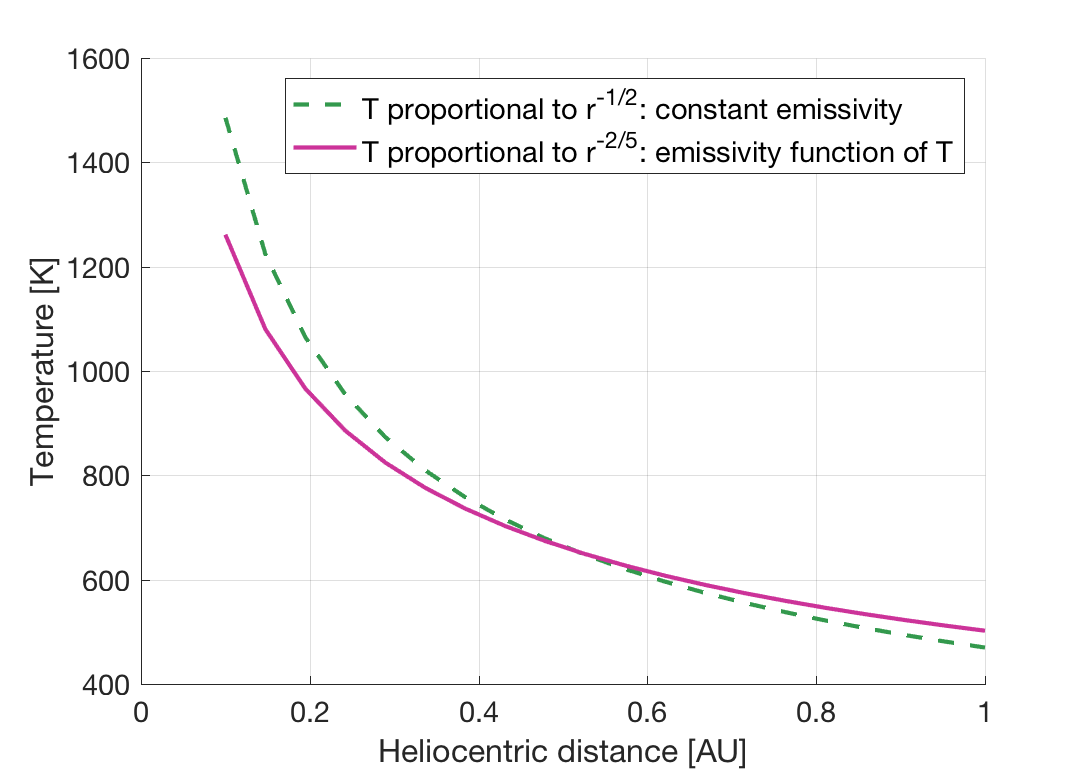}

\selectlanguage{english}%
\end{minipage}

\captionof{figure}{\textsl{Dependence of material's temperature on the heliocentric distance for aluminum (left) and titanium (right), both for constant and variable emissivity.}}

\label{heldist}

\end{figure*}

\subsection{Optical parameters dependence on temperature}

In Eqs. (\ref{eq:Teff}), (\ref{eq:Tact}) and (\ref{eq:Tact-1})
the optical coefficients have been considered constant and not variable
with time or temperature itself. Although reflectivity dependence
on temperature can be neglected, the same doesn't apply for emissivity,
which is directly proportional to the temperature, as suggested by
Parker and Abbott \citep{emiss}, because electrical conductivity
is inversely proportional to $T$. The expression they found for the
total hemispherical emissivity for metals is the following:

\begin{equation}
\begin{array}{c}
{\displaystyle \varepsilon\left(T\right)=7.66\,\sqrt{\frac{T}{\chi\left(T\right)}}+\left[10+8.99\ln\left(\frac{T}{\chi\left(T\right)}\right)\right]}\\
{\displaystyle \times\left(\frac{T}{\chi\left(T\right)}\right)-17.5\left(\frac{T}{\chi\left(T\right)}\right)^{3/2},}
\end{array}\label{eq:zeta-2-1}
\end{equation}
where $\chi\left(T\right)$ is the electrical conductivity in $\Omega^{-1}\:m^{-1}$,
which is defined as the inverse of resistivity $\widetilde{\rho}\left(T\right)$.
As in Fig. \ref{emissivity} it has been shown with experimental results
\citep{radheat} that a very good approximation can be obtained considering
only the first term in Eq. (\ref{eq:zeta-2-1}):

\selectlanguage{american}%
\begin{equation}
\varepsilon\left(T\right)=7.66\,\sqrt{\frac{T}{\chi\left(T\right)}}.\label{eq:zeta}
\end{equation}

\selectlanguage{english}%
The electrical resistivity of most materials changes with temperature.
Usually a linear approximation is used:

\begin{equation}
\widetilde{\rho}\left(T\right)=\widetilde{\rho}_{0}\left[1+\alpha_{t}\left(T-T_{0}\right)\right]\simeq\frac{\widetilde{\rho}_{0}}{T_{0}}\,T,
\end{equation}

\noindent where $\alpha_{t}\simeq1/273\:K^{-1}$ is called the temperature
coefficient of resistivity, $T_{0}$ is a fixed reference temperature
(commonly room temperature) and $\widetilde{\rho}_{0}$ is the resistivity
at temperature $T_{0}$. Values for electrical conductivity, resistivity
and temperature coefficient of various materials can be found in literature.
As $\chi\left(T\right)$ is almost inversely proportional to the temperature,
this means that $\varepsilon\left(T\right)\propto T$: 

\begin{equation}
\varepsilon\left(T\right)=7.66\,\sqrt{T\cdot\widetilde{\rho}\left(T\right)}=7.66\,T\sqrt{\frac{\widetilde{\rho}_{0}}{T_{0}}}.\label{eq:zeta-1-1}
\end{equation}
Hence, by introducing emissivity dependence on temperature in Eq.
(\ref{eq:Tact}), the temperature of the spacecraft varies as $r{}^{-2/5}$
.

\subsection{Results for aluminum and titanium}

In this study the constant optical coefficient were taken from \citep{radheat}.
Experimental data for aluminum are $\rho=0.88$ and $\varepsilon=0.03$,
whereas $\rho=0.22$ and $\varepsilon=0.19$ for titanium. For aluminum
the coefficient in (\ref{eq:zeta}) is $7.52\:K^{-1/2}\:\Omega^{-1/2}\:m^{-1/2}$,
instead for titanium the same given in Eq. (\ref{eq:zeta}) was used.
Considering a reference temperature of $T_{0}=293\:K$, the resistivity
for aluminum and titanium are $\widetilde{\rho}_{0}=2.82\cdot10^{-8}\,\Omega\cdot m$
and $\widetilde{\rho}_{0}=4.2\cdot10^{-7}\,\Omega\cdot m$, respectively.

Dependence of temperature on the heliocentric distance for a case
of constant emissivity and, conversely, when emissivity of the metal
depends on temperature is shown in Fig. \ref{heldist}, and listed
in Table \ref{temp}. It is clear that, considering the temperature
dependence of emissivity (solid curve), the object's temperature increases
more slowly than in the case of constant emissivity (dashed curve),
as the body approaches the Sun \citep{Rom_desorp}. 

\medskip{}

\captionof{table}{\textsl{Temperature dependence of an object on the heliocentric distance for aluminum and titanium considering emissivity constant or function of temperature itself.}}

\begin{singlespace}
\noindent \begin{flushleft}
{\small{}}%
\begin{tabular}{|c|cc|cc|}
\cline{2-5} 
\multicolumn{1}{c|}{} & \multicolumn{2}{c|}{{\small{}$T$ {[}K{]} Al}} & \multicolumn{2}{c|}{{\small{}$T$ {[}K{]} Ti}}\tabularnewline
\cline{1-1} 
{\small{}$r$} & {\small{}$T\propto r^{-1/2}$} & {\small{} $T\propto r^{-2/5}$} & {\small{}$T\propto r^{-1/2}$} & {\small{} $T\propto r^{-2/5}$}\tabularnewline
\cline{1-1} 
{\small{}{[}AU{]}} & {\small{}$\varepsilon=const$} & {\small{}$\varepsilon\left(T\right)$} & {\small{}$\varepsilon=const$} & {\small{}$\varepsilon\left(T\right)$}\tabularnewline
\hline 
{\small{}$0.1$} & {\small{}1476.1} & {\small{}1140.6} & {\small{}1485.7} & {\small{}1261.3}\tabularnewline
{\small{}$0.2$} & {\small{}1043.8} & {\small{}864.4} & {\small{}1050.6} & {\small{}955.9}\tabularnewline
{\small{}$0.3$} & {\small{}852.2} & {\small{}735.0} & {\small{}857.8} & {\small{}812.8}\tabularnewline
{\small{}$0.4$} & {\small{}738.1} & {\small{}655.1} & {\small{}742.9} & {\small{}724.4}\tabularnewline
{\small{}$0.5$} & {\small{}660.1} & {\small{}599.2} & {\small{}664.4} & {\small{}662.6}\tabularnewline
{\small{}$0.6$} & {\small{}602.6} & {\small{}557.0} & {\small{}606.6} & {\small{}616.0}\tabularnewline
{\small{}$0.7$} & {\small{}557.9} & {\small{}523.7} & {\small{}561.6} & {\small{}579.1}\tabularnewline
{\small{}$0.8$} & {\small{}521.9} & {\small{}496.5} & {\small{}525.3} & {\small{}549.0}\tabularnewline
{\small{}$0.9$} & {\small{}492.0} & {\small{}473.6} & {\small{}495.2} & {\small{}523.8}\tabularnewline
{\small{}$1$} & {\small{}466.8} & {\small{}454.1} & {\small{}469.8} & {\small{}502.1}\tabularnewline
\hline 
\end{tabular}
\par\end{flushleft}{\small \par}
\end{singlespace}

\label{temp}

\section{Future development }

Even though aluminum and titanium are the conventional materials for
flight structures, graphite-fiber/polymer-matrix composite materials
are the most probable candidates for the future of aerospace industry.
Carbon-fiber composite materials are easier to shape and lighter than
metals, with much higher strength to density ratio \citep{TechSC}.
Significant weight savings (25 to 50 percent) could be achieved through
the use of polymer-matrix composites. However, a great engineering
effort is required to establish confidence in their use. Our analysis
could be easily extended to these materials, once their optical properties
are known. Nevertheless, due to the unicity of each composite material,
experimental values may be required.

\section{Conclusions}

Within the standard approach the reflectivity and emissivity of materials
used in aerospace industry are conserved as constant values. As a
result the temperature of the material increases when the spacecraft
approaches to the sun as $T\sim r{}^{-1/2}$, where $r$ is the heliocentric
distance. In our approach we consider the temperature dependence of
the electro-optical parameters of materials that have implicit dependence
on the temperature through the temperature dependence of the electrical
conductivity of metals. It is shown that the temperature dependence
of the emissivity and conductivity leads to the dependence of temperature
for the materials used in aerospace industry approximately as $T\sim r{}^{-2/5}$
with the heliocentric distance. The proposed study compares the temperature
dependence on heliocentric distance for metals (aluminum and titanium
in particular) when their optical properties are considered constant
or function of temperature. For a near-Sun mission it is of crucial
importance to understand as much as possible the behavior of the spacecraft
materials, as the greatest constraint will be the maximum temperature
at which it can be fully functional and operative. 

In our study the temperature reached at 0.1 AU with variable emissivity
is lower than when one assumes that the emissivity is constant, both
for Al and Ti. In other words, the spacecraft could approach closer
to the Sun than it was assumed before based on the consideration that
emissivity of materials used in aerospace industry does not depend
on temperature. However, note that the titanium graph shows an intersection
between the two curves. This indicates that there is no common behavior
and the study must be carried out with a complete set of the chosen
material's electro-optical parameters.

\section*{Acknowledgements }

This research was supported by PSC CUNY Grant: award \# 68298-0046.

\bibliographystyle{unsrtnat}


\begin{thebibliography}{99}
	
\bibitem{Rom_desorp}	R. Ya. Kezerashvili. Space exploration with a solar sail coated by materials that undergo thermal desorption. \emph{Acta Astronaut.}, 117:231-237, 2015.
	
\bibitem{MatSP}	M. Peters and C. Leyens. Aerospace and space materials. \emph{Materials Science and Engineering}.

\bibitem{heatmass} Y. A. \c{C}engel and A. J. Ghajar. \emph{Heat and Mass Transfer - Fundamentals and Applications.} McGraw-Hill Education, 5th edition, 2015.

\bibitem{ClarkeAstr} A. E. Roy and D. Clarke. Astronomy: Principles and practice. \emph{Taylor and Francis, Fourth Edition}, 2003.

\bibitem{materials} R. Ya. Kezerashvili. Solar sail: Materials and space environmental effects.\emph{in the book: M. Macdonald (Ed.), Advances in Solar Sailing, Springer Praxis Books, Berlin, Heidelberg}, pages 573?592, 2014.

\bibitem{radheat} M. F. Modest. \emph{Radiative Heat Transfer}. McGraw- Hill, 1993.

\bibitem{emiss} W. J. Parker and G. L. Abbott. Theoretical and ex- perimental studies of the total emittance of metals. \emph{Symposium on Thermal Radiation of Solids, NASA SP-55}, pages 11?28, 1965.

\bibitem{TechSC} National Research Council L. J. Adams. Technol- ogy for small spacecraft. \emph{The Aeronautics and Space Engineering Board}, 1994.

\end{thebibliography}

\end{multicols}{}
\end{document}